\documentclass[twocolumn,showpacs,preprintnumbers,showkeys,amsmath,amssymb,prl]{revtex4}

\usepackage{graphicx}
\usepackage{dcolumn}
\usepackage{bm}

\makeatletter

\newcommand{\Rmnum}[1]{\expandafter\@slowromancap\romannumeral #1@}

\makeatother 	
	
\begin{document}

\title{Polymer dynamics, fluorescence correlation spectroscopy, and the limits of optical resolution}

\author{J\"org Enderlein}
 \email{enderlein@physik3.gwdg.de}
 \affiliation{\Rmnum{3}. Institute of Physics, Georg August University, 37077 G\"ottingen, Germany.}

\date{\today}

\begin{abstract}

In recent years, fluorescence correlation spectroscopy has been increasingly applied for the study of polymer dynamics on the nanometer scale. The core idea is to extract, from a measured autocorrelation curve, an effective mean-square displacement function that contains information about  the underlying conformational dynamics. The paper presents a fundamental study of the applicability of fluorescence correlation spectroscopy for the investigation of nanoscale conformational  and diffusional dynamics. We find that fluorescence correlation spectroscopy cannot reliably elucidate processes on length scales much smaller than the resolution limit of the optics used and that its improper use can yield spurious results for the observed dynamics.

\end{abstract}



\maketitle

The advent of single molecule fluorescence techniques has allowed for the direct visualization and measurement of the conformation and dynamics of individual polymer molecules, see e.g. \cite{Quake1997, LeDuc1999,  Cohen2007, Woll2009, McHale2009}. However, this single-molecule imaging approach is restricted to the study of polymers which are larger in size than the diffraction limit of the imaging optics. Fluorescence correlation spectroscopy (FCS) is a technique that can purportedly resolve dynamics below the diffraction limit. In a recent paper, Shusterman et al. \cite{Shusterman2004} introduced an analysis method for FCS data, which they used to analyze the nanoscale conformational dynamics of single- and double-stranded DNA, reporting Zimm instead of Rouse dynamics for the studied molecules. This finding has spurred considerable theoretical work in an effort to explain this unexpected result \cite{Winkler2007, Hinczewski2009, Hinczewski2011}. Meanwhile, the analysis method of Shusterman et al.~has found widespread application in FCS-based studies of polymer dynamics Ref.~\cite{Bernheim-Groswasser2006, Lisy2006, Tothova2007, Shusterman2008, Wocjan2009}. 

Here, we present a fundamental analysis of the capability of FCS  to elucidate polymer dynamics on the nanometer scale. If FCS is indeed capable of measuring the conformational dynamics of small molecules close to and beyond the diffraction limit of the used optics, this would be of tremendous interest. Not only could one resolve the question of whether Rouse or Zimm dynamics is the appropriate model for the dynamics of disordered polymers, but one could also experimentally study problems such as protein folding and unfolding, or the nanoscale diffusion of molecules within lipid membranes.  

Thus, the core question is how well FCS can elucidate nanoscale conformational/diffusional dynamics. Let us start with a simple one-dimensional system, namely the one-dimensional free diffusion, with diffusion coefficient $D_0$, of a molecule along the $x$-axis within the confined region $-a \leq x \leq a$, where the confinement itself diffuses with a much slower diffusion coefficient $D$ through the focus of a FCS system. For the diffusion within the confined region, Green's function is given in a standard way as an expansion over eigenfunctions 

\begin{eqnarray}
G(x,x^\prime,t) = \sum_{j=0}^\infty \phi_j(x) \phi_j(x^\prime) \exp(-\omega_j D_0 t)
\end{eqnarray}

\noindent where the eigenfunctions $\phi_{j}(x)$ are non-zero only for $|x| \leq a$, and where they are $\phi_{2j}(x) = \cos(j\pi x/a)$ for even indices, and $\phi_{2j+1}(x) = \sin((j+1/2) \pi x/a)$ for odd indices. The characteristic frequencies are $\omega_j = (j \pi/2a)^2$. 

Let us further assume that fluorescence is excited by focusing a plane wave through a lens with numerical aperture NA into a medium with refractive index $n$. The resulting intensity distribution is given by

\begin{eqnarray}
U(x) = \left[ \frac{\sin(k_\text{max} x)} {k_\text{max} x} \right]^2
\label{eq:focus}
\end{eqnarray}

\noindent where $k_\text{max} = 2 \pi \text{NA}/\lambda$. For 500~nm wavelength light, a numerical aperture of 1.2, and a refractive index of 1.33 (water), one can estimate the focus width and thus spatial resolution of a corresponding scanning microscope to be of ca. 200 nm. The best fit of a Gaussian distribution to the actual intensity distribution, Eq.~(\ref{eq:focus}), is found to have a Gaussian width of $2 \sigma = 152$~nm. Assuming that the excited fluorescence is detected uniformly over the whole excitation region, then $U(x)$ is directly proportional to the molecule detection function (MDF) of the system, which gives the probability density of detecting a photon from a molecule at position $x$. The autocorrelation function (ACF) of an FCS measurement is then given as the multiple integral over the product of the probability to detect a photon from a molecule at some initial position, the probability density that it diffuses from this position to a final position within time $t$ (given by Green's function), and the probability to detect a photon from a molecule at this final position. Thus, the short-time behavior of the ACF can be written as 

\begin{eqnarray}
g(t) = \iiint dx\, dx^\prime \,dy \,U(x+y) G(x,x^\prime,t) U(x^\prime+y)
\label{eq:ACF}
\end{eqnarray}

\noindent where $y$ is the center position of the confined region with respect to the excitation focus. The integrations over $x$ and $x^\prime$ run from $-a$ to $a$, and the integration over $y$ from $-\infty$ to $\infty$. In Eq.~(\ref{eq:ACF}), we have neglected all constant pre-factors related to concentration, overall detection efficiency, etc. "Short-time behavior" means that we consider here only a time range where the fast confined diffusion dominates the ACF, before the slow diffusion of the confined region itself plays any role. Integration over $y$ takes into account that the confined region can be at any position with respect  to the excitation focus. Eq.~(\ref{eq:ACF}) can be rewritten by switching to Fourier space, representing all functions by their Fourier transforms, e.g.

\begin{eqnarray}
U(x) = \int\limits_{-\infty}^{\infty} \frac{dk}{2\pi} \, \tilde{U}(k) \exp(i k x)
\end{eqnarray}

\noindent After carrying out several integrations, one finds the compact expression

\begin{eqnarray}
g(t) =\sum_{j=0}^\infty a_j \exp(-\omega_j D_0 t)
\label{eq:gsum}
\end{eqnarray}

\noindent where the amplitudes $a_j$ are given by

\begin{eqnarray}
a_j = \int\limits_{-\infty}^{\infty} \frac{dk}{2\pi} \left|\tilde{U}(k)\tilde{\phi}_j(k) \right|^2
\label{eq:gamp}
\end{eqnarray}

\begin{figure}
\centering\includegraphics[keepaspectratio, width=8cm]{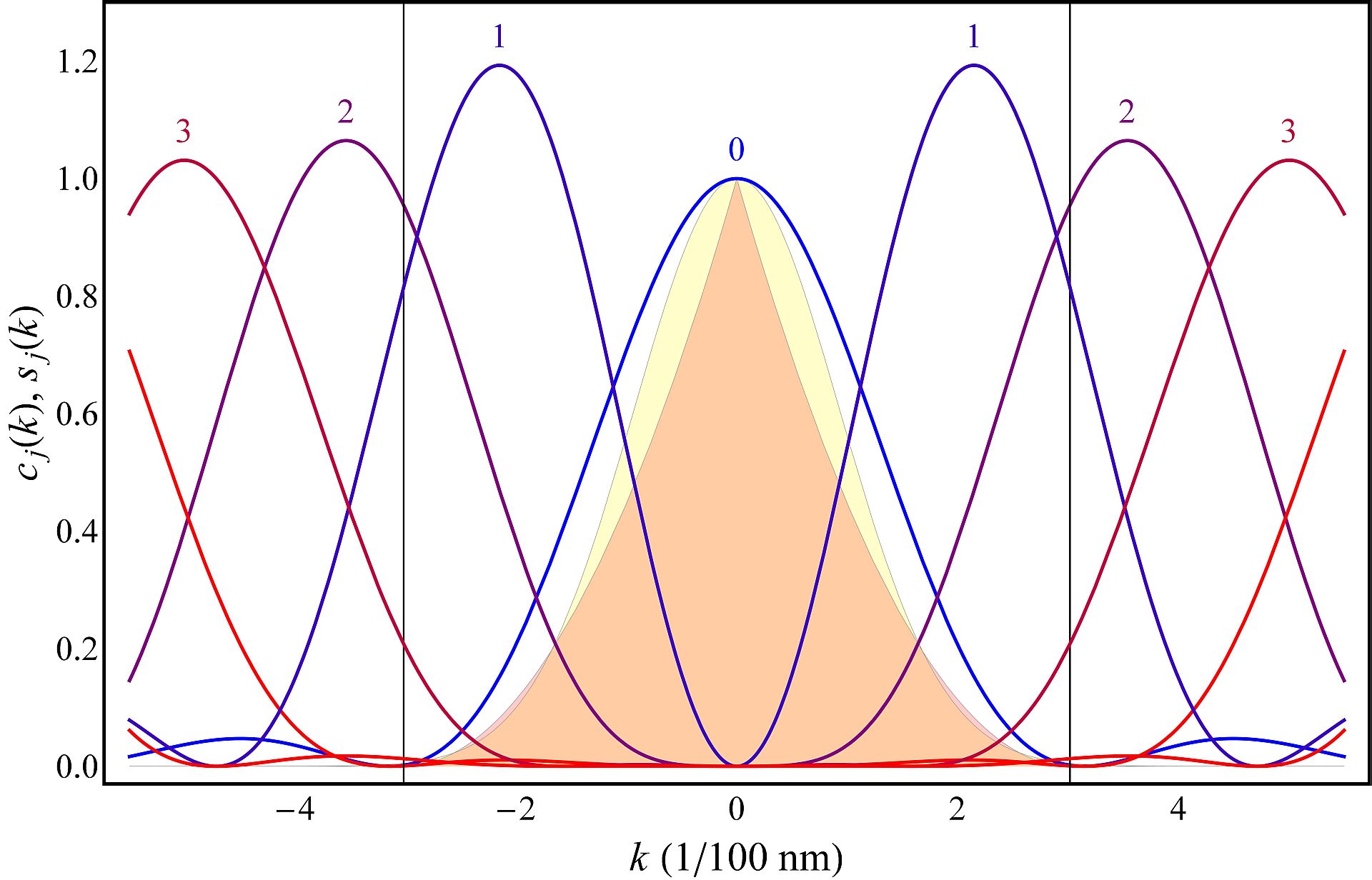}
\caption{\label{fig:fourier} Fourier power spectra of the MDF and the eigenfunctions $\phi_j$ entering the Green's function for confined diffusion. The curves $j=0,1,...$, represent the power spectra $|\tilde{\phi}_j(k)|^2$ for $a=50$~nm. The triangular shaded region is the power spectrum $|\tilde{U}(k)|^2$, and the yellow shaded region that of the Gaussian approximation of $U(x)$. Vertical lines delimit the finite support of the optical transfer function. The position of the maxima of the functions $|\tilde{\phi}_j(k)|^2$ scales with the inverse value of $a$. Thus, the smaller the confinement region, the more the maxima will be shifted away from the  Fourier modes that can be probed by the MDF.}
\end{figure}

\noindent Eqs.~(\ref{eq:gsum}) and (\ref{eq:gamp}) are particularly suited for analyzing the impact of the optical resolution of the measurement setup on the temporal behavior of an ACF. As can be seen, the amplitude for the $j$th exponential function in Eq.~(\ref{eq:gsum}) is given by the weighted integral over the power spectrum of the $j$th eigenfunction $\phi_j$, where the weight function is the power spectrum of the MDF. However, for increasing numbers of $j$, the power spectrum of $\phi_j$ is shifted to higher and higher spatial frequencies $k$, whereas the filter $\tilde{U}(k)$ is non-zero only for $k<k_\text{max}$. A visualization of this fact is presented in Fig.~\ref{fig:fourier}, showing the power spectra for the first few eigenfunctions together with the power spectrum of the MDF (both for the exact MDF as well for its Gaussian approximation). The calculations were done for $a=50$~nm, thus confining diffusion to a region of 100 nm. As can be seen, there is substantial overlap between $|\tilde{U}(k)|^2$ and $|\tilde{\phi}_j(k)|^2$ for the first three eigenfunctions only -- all the information contained in the higher order eigenfunctions which correspond to faster time scales (larger values of $\omega_j$) will not be reflected in the ACF.

The core idea of Shusterman et al.~\cite{Shusterman2004} is to compare a measured ACF with the model ACF for free-diffusion in the Gaussian approximation of the MDF. For the one-dimensional case considered here, such a model ACF is given, up to a constant pre-factor, by the simple expression 

\begin{eqnarray}
g(t,\sigma) = \frac{1}{\sqrt{1+\langle x(t)^2 \rangle/2\sigma^2}}
\label{eq:gmodel}
\end{eqnarray}

\noindent where $\langle x(t)^2 \rangle= 2 D_0 t$ is the mean square displacement of a molecule within time $t$, and $\sigma$ the variance of the Gaussian that is used for approximating the MDF. Under the assumption that this form of the ACF will also be valid for more complex diffusion behavior, Eq.~(\ref{eq:gmodel}) can be inverted to yield the alleged relationship

\begin{eqnarray}
\langle x(t)^2 \rangle \propto \frac{g_0^2}{g^2(t)} - 1
\label{eq:x2}
\end{eqnarray}

\begin{figure}
\centering\includegraphics[keepaspectratio, width=8cm]{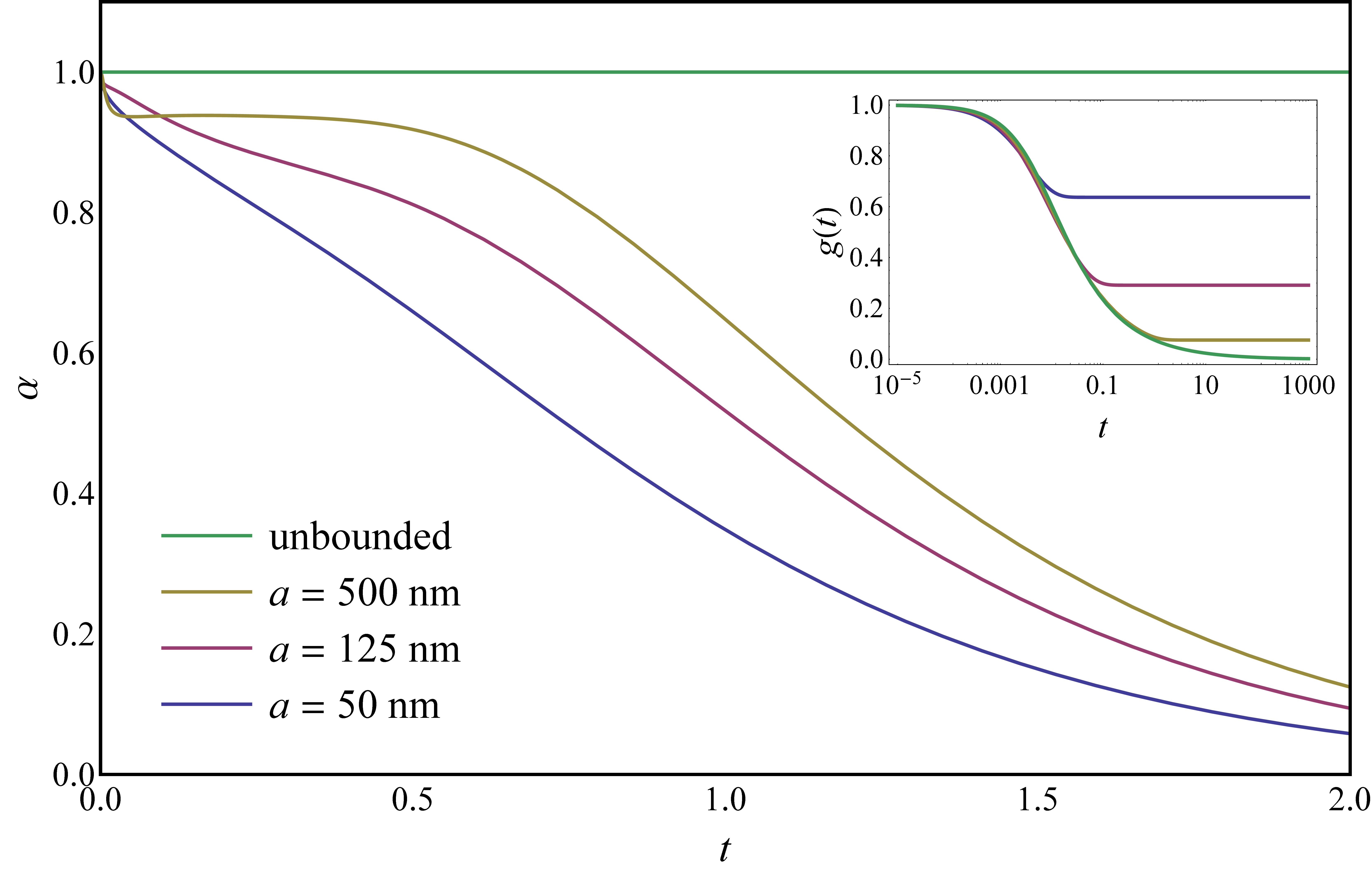}
\caption{\label{fig:1d} Behavior of the function $\alpha(t)$ for different values of the confinement parameter $a$ and for $D_0 = 1 \; \mu$m$^2/$time unit. The inset shows the different ACFs for the selected values of $a$.}
\end{figure}

\noindent For free diffusion, the right hand side of Eq.~(\ref{eq:x2}) is a linear function of time. For confined diffusion, one would still hope to see the linear behavior on short time scales. Let us define a quantity $\alpha(t)$ as

\begin{eqnarray}
\alpha(t) = \frac{d \ln \langle x(t)^2 \rangle}{d \ln t}
\label{eq:alpha1}
\end{eqnarray}

\noindent which defines the local exponent with which $\langle x(t)^2 \rangle$ increases with time $t$. This function is time-independent and equal to one for free diffusion, and one also expects that it will be close to one on sufficiently short time scales for confined diffusion. Fig.~\ref{fig:1d} shows the behavior of $\alpha(t)$ for several selected values of the confinement parameter $a$. Indeed, $\alpha(t)$ converges to unity in the limit of zero  time, but this is true for \emph{any} well-behaved function $g(t)$! However, not much can be learned from $\alpha(t)$ for intermediate  time values. This is due to the fact that with increasing confinement (decreasing value of $a$) the correlation function captures less and less of the diffusion dynamics due to the finite support of the MDF in Fourier space. It is important to notice that we consider here only the limiting case of negligible diffusion of the confined area as compared to the diffusion of a molecule within the confined area. At large  times, the diffusion of the whole confinement region will start to show up in the ACF, thus leading again to values of $\alpha(t)$ close to unity. In the intermediate region, however, the curves in Fig.~\ref{fig:1d} show that one can extract any fractional power dependence of $\langle x(t)^2 \rangle$ on time $t$. 

To consider a more realistic scenario, and to prove that the behavior of $\alpha(t)$ is not an artifact of the simple one-dimensional model system considered so far, let us study the rapid diffusion of a particle in three dimensions within an isotropic harmonic potential, $V(\mathbf{R}) =  k_B T R^2/2 \kappa$, where $k_B$ is Boltzmann's constant, $T$ is temperature, and $\kappa$ is a constant having units of length squared which determines the steepness of the potential. The coordinate vector $\mathbf{R}$ refers to the center position $\mathbf{r}_c$ of the potential, $\mathbf{R} = \mathbf{r}-\mathbf{r}_c$, whereas the center position $\mathbf{r}_c$ is assumed to diffuse freely through space (slow diffusion). Denoting the diffusion coefficient for the rapid diffusion again by $D_0$, Green's function for the diffusion within the harmonic potential is given by 

\begin{eqnarray}
G(\mathbf{R},\mathbf{R}_0,t) = \frac{1} {\left[ 2 \pi \kappa \left( 1-s^2 \right) \right]^{3/2}}  \exp\left[ -\frac{\left\vert\mathbf{R} - \mathbf{R}_0 s \right\vert^2}{2 \kappa \left( 1-s^2 \right) } \right]
\label{eq:square}
\end{eqnarray}

\noindent where we have used the abbreviation $s = e^{-D_0 t/\kappa}$. Furthermore, for the slow diffusion of the center position $\mathbf{r}_c$ with diffusion constant $D < D_0$, we have the Green's function for the free diffusion equation

\begin{eqnarray}
G_c(\mathbf{r}_c-\mathbf{r}_{c,0},t) = \frac{1}{\left(4 \pi D t\right)^{3/2}} \exp\left[-\frac{\left\vert\mathbf{r}_c - \mathbf{r}_{c,0} \right\vert^2}{4 D t} \right]
\label{eq:free}
\end{eqnarray}

\noindent In most FCS publications, and also in Ref.\cite{Shusterman2004,Lisy2006}, the three-dimensional MDF, $U(\mathbf{r})$, of a confocal fluorescence microscope is approximated by an axi-symmetric Gaussian distribution with half axes $\sigma$ and $\zeta \sigma$. For calculating the final ACF, we still need the equilibrium probability $p_0(\mathbf{R}_0)$ of finding a molecule at some initial position $\mathbf{R}_0$ within the harmonic potential, which is found by letting time $t$ in Eq.~(\ref{eq:square}) approach infinity. The ACF is then equal to the multiple integral of the product of the initial probability $p_0(\mathbf{R}_0)$, the probability density $G(\mathbf{R},\mathbf{R}_0,t)$ that the molecule moves from initial position $\mathbf{R}_0$ to position $\mathbf{R}$ within the harmonic potential during time $t$, the probability density $G_c(\mathbf{r}_c-\mathbf{r}_{c,0},t)$ that the center position diffuses from $\mathbf{r}_{c,0}$ to position $\mathbf{r}_c$ within the same time $t$, and the probabilities $U(\mathbf{r}_0)$ and $U(\mathbf{r})$ to detect a photon at the initial and final position. After carrying out all twelve integrations over all possible initial and final positions of molecule and potential center, one finally arrives at the analytic expression (up to a constant pre-factor)

\begin{eqnarray}
g(t) = f(\sigma,t)^2 f(\zeta \sigma,t)
\label{eq:g3d}
\end{eqnarray}

\noindent where

\begin{eqnarray}
f(q,t) = \left[ 1+ \frac{D t + \kappa \left(1-s\right)}{q^2}  \right]^{-1/2}
\label{eq:f}
\end{eqnarray}

\begin{figure}
\centering\includegraphics[keepaspectratio, width=8cm]{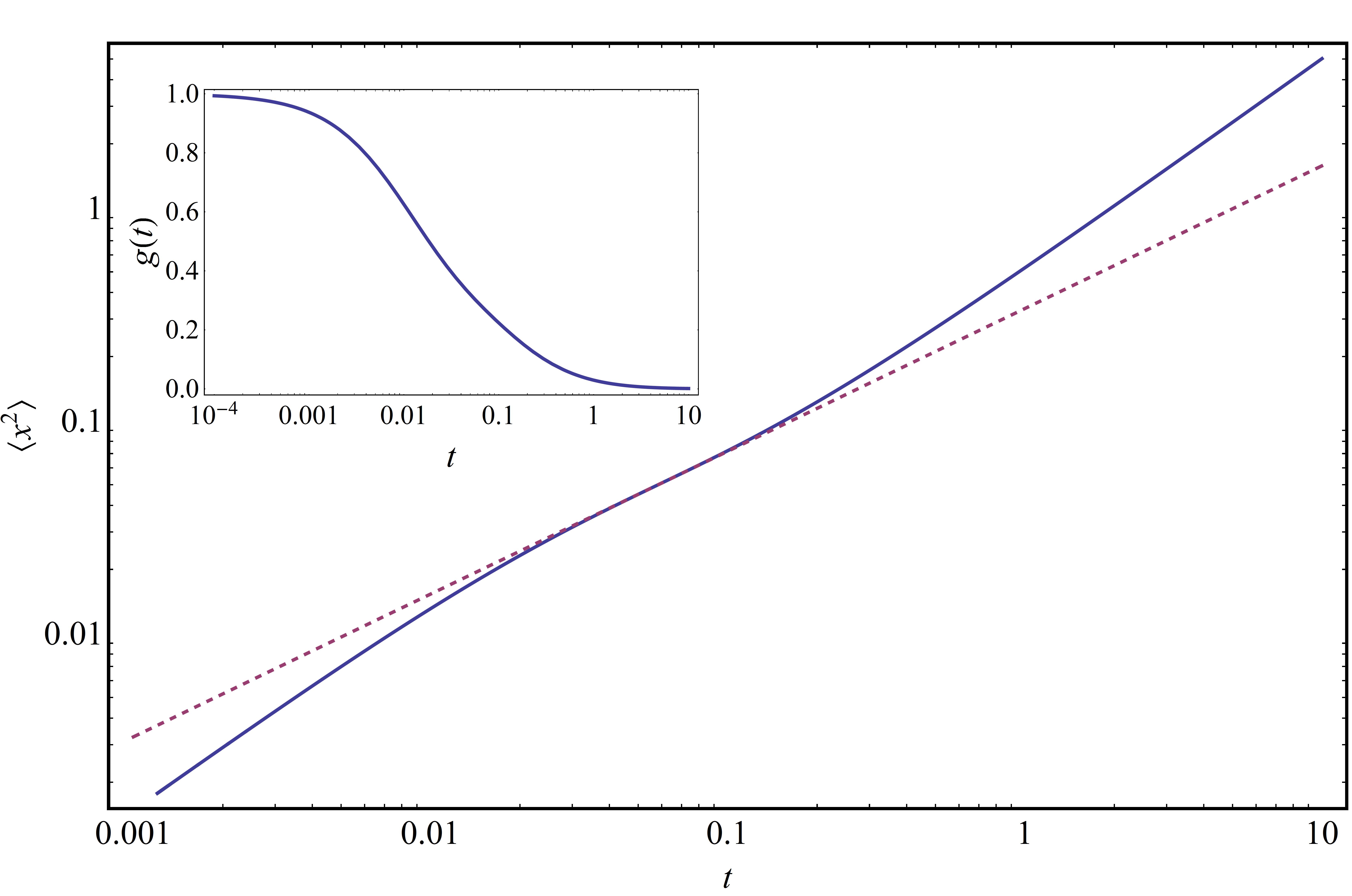}
\caption{\label{fig:g3d} Mean-square displacement $\langle x^2 \rangle$ as a function of time determined by inverting Eq.~(\ref{eq:g3free}) for the parameter values given in the text. Inset shows the corresponding ACF as calculated from eqs.\ref{eq:g3d} and \ref{eq:f}. The dashed line is a linear fit of $\ln t$ to $\ln \langle x^2 \rangle$ at the point of minimum slope.}
\end{figure}

\noindent As an example, the correlation function is calculated for the  following parameters: the shape of the MDF is determined by $\sigma=150$~nm and $\zeta = 5$, the fast diffusion coefficient within the potential is $D_0 = 1 \, \mu$m$^2/$time unit, the diffusion coefficient $D$ of the diffusion of the center position of the harmonic potential is set equal to 0.5~$\mu$m$^2/$time unit, and $\kappa$ is chosen to be $(150~\text{nm})^2$. Similar to the one-dimensional case, one implicitly defines an effective mean-square displacement function $\langle x^2 \rangle$ by the relation \cite{Shusterman2004}

\begin{eqnarray}
g(t) = \frac{1}{(1+\langle x^2 \rangle/\sigma^2) \sqrt{1+\langle x^2 \rangle/(\zeta \sigma)^2}}
\label{eq:g3free}
\end{eqnarray}

\begin{figure}
\centering\includegraphics[keepaspectratio, width=8cm]{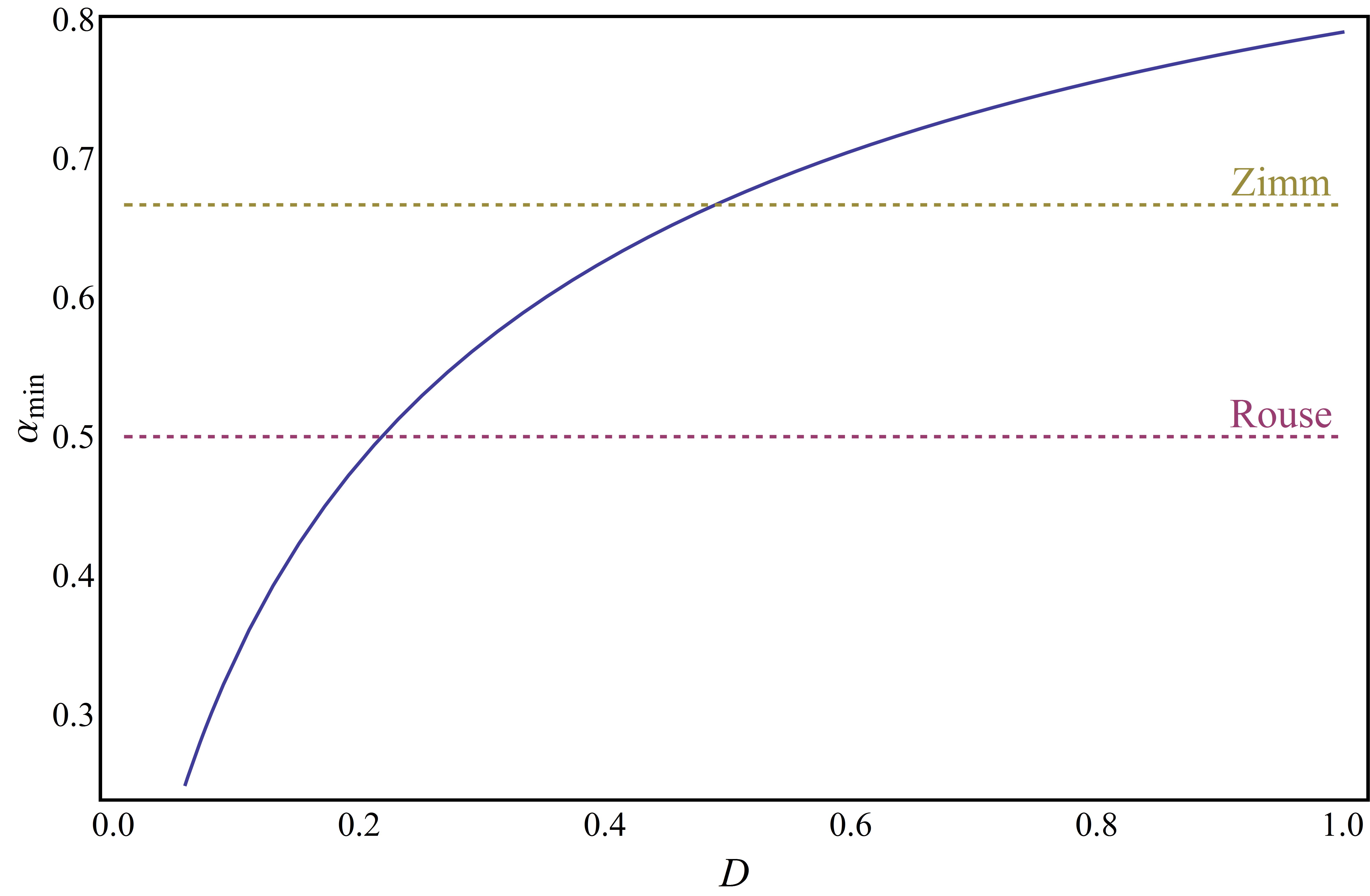}
\caption{\label{fig:alpha3d} Minimum value of $\alpha(t)$ as a function of the ratio $D/D_0$. Shown are also the expected values of $\alpha$ for Zimm and Rouse dynamics.}
\end{figure}

\noindent Fig.~\ref{fig:g3d} shows the behavior of $\langle x^2 \rangle$ as a function of time, together with the corresponding shape of the ACF (inset). The broken line represents a linear fit of $\ln t$ to the curve of $\ln \langle x^2 \rangle$ at the point of minimum slope. For the chosen parameters, this slope value is ca. 2/3, mimicking a diffusion process originating from a Zimm dynamics scenario. Let us define again a characteristic function $\alpha(t)$ as in Eq.~(\ref{eq:alpha1}). Generally, $\alpha(t)$ will approach unity for very small and very large values of  time $t$, but will have a minimum at intermediate time values. Fig.~\ref{fig:alpha3d} shows the minimum value of $\alpha(t)$ as a function of the ratio $D/D_0$. As can be seen, depending on this value, the function $\langle x^2 \rangle$ can show intermediate power-law behavior with \emph{any} exponent, although the underlying process is simple diffusion within a harmonic potential.\\

In summary, FCS measurements are governed by the same spatial resolution limit as any linear optical imaging system, and the smaller the length scale of a diffusion process the less information FCS can capture about it. Moreover, defining and analyzing an effective mean-square displacement function $\langle x^2 \rangle$ as first proposed in Ref.~\cite{Shusterman2004} can produce any power-law behavior with little connection to the actual underlying physical process. However, as was shown when analyzing the interplay between the Fourier spectra of the eigenfunctions entering the Green's function of the diffusion equation and the Fourier spectrum of the MDF (Fig.~\ref{fig:fourier}), an FCS experiment cannot capture much of the dynamics on length scales that are outside of the finite Fourier support of the MDF of the optical system. Thus, our analysis highlights the general limitation of FCS for elucidating processes on length scales below the resolution limit of the optics used in an FCS experiment. 

\begin{acknowledgments}
Financial support by the Deutsche Forschungsgemeinschaft is gratefully acknowledged (SFB 755, project A5). I thank Ingo Gregor and Christoph Pieper for many fruitful discussions and for critically reading the manuscript. Linguistic help from Chris Battle is very much appreciated.
\end{acknowledgments}


\end{document}